**Title:** Comparing causal estimands from sequential nested versus single point target trials: A simulation study

**Authors:** Catherine Wiener[1,2]*, Chase D. Latour[2,3]*, Kathleen Hurwitz[2], Xiaojuan Li,[4] Catherine Lesko[5], Alexander Breskin[6], M. Alan Brookhart[2,7]
* These authors contributed equally to this work.

**Affiliations**:
[1] Department of Epidemiology, University of North Carolina at Chapel Hill, Chapel Hill, NC USA
[2] Target RWE, Durham, NC USA
[3] Department of Biostatistics, Epidemiology, and Informatics, Perelman School of Medicine, University of Pennsylvania, Philadelphia, PA USA
[4] Department of Population Medicine, Harvard Medical School and Harvard Pilgrim Health Care Institute, Boston, MA USA
[5] Department of Epidemiology, Bloomberg School of Public Health, Johns Hopkins University, Baltimore, Maryland
[6] Regeneron Pharmaceuticals, Tarrytown, NY USA
[7] Department of Population Health Sciences, Duke University, Durham, NC USA

**Corresponding Author:**
Catherine Wiener
Target RWE
212 W Main Street, Ste 308
Durham, NC 27701
catiewien@gmail.com

**Data Availability:** Data for this study were simulated using R version 4.5.0 (R Foundation for Statistical Computing, Vienna, Austria). Code is available on GitHub: [https://github.com/ktwiener/publications/tree/main/snt_emulation].

**Conflicts of Interest:** CDL received payment from TargetRWE for this work and has received payment from Regeneron Pharmaceuticals and Amgen for unrelated work. CW has received payment from TargetRWE for this work and unrelated work. AB is a an employee and shareholder of Regeneron Pharmaceuticals. KEH is an employee of and owns equity in Target RWE. MAB serves on scientific advisory committees for the American Association for Allergy Asthma and Immunology, Amgen, Brigham and Women's Hospital, Gilead Sciences, Intercept, National Institute of Diabetes and Digestive and Kidney Diseases, Target RWE, and Regeneron; he also has equity with AccompanyHealth, Target RWE, and VitriVax.

**Acknowledgements:** We would like to thank Paul Zivich, PhD for his suggestions to our simulation approach.

## Abstract

Sequential nested trial (SNT) emulation is a powerful approach for maximizing precision and avoiding time-related biases. However, there exists little discussion about the implied causal estimands in comparison to a real-world single point trial. We used Monte Carlo simulation to compare treatment effect estimates from an SNT emulation that re-indexed patients annually and a SNT emulation with a treatment decision design to the estimates from a single point trial. We generated 5,000 cohorts of 5,000 people with 3 years of follow-up. For the single point trial, patients were randomized to initiate or not initiate treatment at Visit 1. For the SNT emulations, simulated patients could contribute up to two index dates. When disease severity did not modify the treatment effect, both SNT approaches returned treatment effect estimates identical to the single point trial. In the presence of treatment effect modification by disease severity, both SNT approaches returned treatment effect estimates that diverged from the single point trial even after confounding-adjustment. These findings underscore the difficulties of interpreting causal estimands from a SNT emulation: the target population does not correspond to a single time point trial. Such implications are important for communicating study results for evidence-based decision-making.

## 1. Introduction

Target trial emulation is a framework for designing observational studies to emulate a pragmatic, hypothetical randomized trial (the *target trial*).[1–8] This framework helps researchers articulate an unambiguous research question and avoid time-related biases. Multiple approaches exist to emulate a specified target trial.[1,5,9–11] One approach involves emulating a series of nested target trials, or *sequential nested trials* (SNTs). In a hypothetical SNT, individuals are indexed into the study the first time they meet eligibility criteria, and are randomized to initiate treatment (or not) at that time.[4,10–17] Individuals not initially assigned to treatment may be subsequently re-randomized with new, additional time origins, or indexes, until randomized to treatment or the end of follow-up.

A SNT emulation follows the same design as the hypothetical SNT but is implemented in observational data such that treatment is observed at each eligible index rather than randomized. Individuals contribute indexes until they are observed to have initiated treatment. There are multiple ways to index individuals in a SNT emulation: at regular intervals (e.g., each week or year after study entry)[11,15] or at treatment decision points (e.g., encounters where laboratory values are above a clinical threshold).[18,19] Regardless of indexing approach, SNTs leverage longitudinal data and multiple treatment opportunities to improve statistical efficiency by increasing the number of analyzed indexes.

Despite these statistical advantages, causal estimands from SNT emulations can be difficult to interpret when compared to emulation approaches that only include a person's first eligible index. In SNT emulations, the implicit target population consists of a dynamic set of time origins, capturing individuals at multiple points in their disease course. As a result, common causal estimands like the average treatment effect (ATE) or the average treatment effect in the treated (ATT) may no longer correspond to effects in a well-defined population of individuals, but rather to an average over a mixture of person-time units at varying disease stages.

Our objective was to explore how the choice of indexing approach and distribution of time-varying effect modifiers impacts causal effect estimates produced by SNT emulation versus a single point trial (SPT). If effect measure modifiers change over time, these decisions could cause the effect estimate we derive from an SNT emulation to deviate substantially from that we would expect in a conventional SPT. Given that much of our decision-making has historically relied on effect estimates from SPTs, such deviations could meaningfully complicate the interpretation and application of results. We compare estimates of the ATE and the ATT in three studies: (1) a SPT, (2) a calendar-based SNT emulation wherein non-initiators contributed indexes at 1-year intervals (hereafter, the eSNT-CAL), and (3) a treatment decision based SNT emulation wherein non-initiators contributed time origins at treatment decision points (hereafter, the eSNT-TD).

## 2. Methods

### *2.1. Motivating Study*

We motivated our simulation with a sequential nested trial emulation conducted by Brookhart et al. (2024) to answer the following clinical question: What is the effect of initiating versus not initiating obeticholic acid (OCA) on the two-year risk of a composite outcome of death, liver transplant, or hospitalization for hepatic decompensation among patients with primary biliary cholangitis (PBC)?[19] PBC is a moderately rare, chronic, cholestatic, autoimmune disease that frequently progresses to cirrhosis and liver failure, requiring liver transplant.[20] PBC can vary in severity, which influences the probability of the outcome, and, at the time of the study, OCA initiation was rare in real-world populations. SNT emulation allowed for more treated indexes to be included in the study, thereby improving precision.

*2.2. Single Point Trial (SPT)*

The simulated SPT represented the ideal randomized trial to answer the motivating clinical question and defined our target population (**Table 1**). We considered a SPT with two inclusion criteria:
  (1) diagnosis of PBC without prior OCA use, or
  (2) a "treatment decision point" – a clinical visit with a laboratory test value indicating OCA eligibility (e.g., total bilirubin above the upper limit of normal).[19]

Patients were indexed into the study at the first visit where they met eligibility criteria and were randomized to initiate or never initiate OCA (Supplemental Figure 1 (S1)). We followed patients from randomization until the first of the composite outcome or the end of follow-up (2 years). See Section 2.4 for details on the data generation for the SPT.

We focused on the two-year per-protocol treatment effect on the composite outcome: the effect had everyone initiated and remained adherent to OCA versus had no one ever initiated OCA. We estimated three parameters: the crude risk ratio (RR), the RR for the ATE, and the RR for the ATT (Table S1).

*2.3. Sequential Nested Trials (SNTs)*

The sequential nested target trial repeatedly applies the single point target trial over multiple indexes and is the trial targeted by the SNT emulations in observational data. Below, we describe both the target trials and their emulations for the calendar-based SNT emulation and the treatment decision-based SNT emulation. Only data for the SNT *emulations* were simulated.

*2.3.1. Sequential Nested Target Trials*

We specified two sequential nested target trials, mirroring the eligibility criteria from the SPT (**Table 1**). In both, patients would be initially indexed into the study at their first encounter (Visit 1) that met inclusion criteria. Individuals randomized to not initiate OCA could then be re-randomized at a subsequent encounter one year later (Visit 2). For the calendar-based sequential nested target trial (corresponding to inclusion criteria [1]), all non-OCA patients who had not experienced the study outcome would be re-randomized at Visit 2. For the treatment decision-based sequential nested target trial (corresponding to inclusion criteria [2]), patients would be re-randomized only if the second visit met the treatment decision criterion.

Like the SPT, in observational data, we would follow indexes until the composite outcome, the end of follow-up (2 years), or, for untreated indexes, time of OCA initiation.

*2.3.2. Sequential Nested Trial Emulations (eSNT-CAL and eSNT-TD)*

We simulated both the calendar-based SNT emulation (eSNT-CAL) and the treatment decision-based SNT emulation (eSNT-TD) (Table 1). For the emulations, patients were initially indexed into the studies at their first encounter meeting inclusion criteria. Treatment was determined based upon OCA initiation at that encounter. Subsequent OCA initiation could occur at a visit one year later. For eSNT-CAL (inclusion criteria [1]), all individuals who did not initiate OCA at their first index contributed a second index event one year later. For the eSNT-TD (inclusion criteria [2]), individuals who did not initiate OCA at their first index contributed index events 1-year post-initial index only if those encounters met treatment decision criteria.

We conducted per-protocol analyses on the index-level (Table S1). OCA indexes were followed until the composite outcome or the end of follow-up, whichever occurred first. Non-OCA indexes were followed until OCA initiation, the composite outcome, or the end of follow-up, whichever occurred first.

## 2.4. Simulating Patient Data

*2.4.1 Superpopulation and Potential Outcomes*

We generated a superpopulation (N=100,000,000) for each scenario (Section 2.5) based on the data generating mechanism implied by the directed acyclic graphs in Figure 1.

For each simulated individual in the superpopulation, we simulated 3 follow-up visits at regular, 1-year intervals and generated PBC severity (low versus high) for Visits 1-3 and an indicator as to whether Visit 2 met criteria for a treatment decision. We generated discrete potential outcomes at times 2-4 under OCA treatment and no treatment, conditional on PBC severity at the prior visit (Figure S2). To control the two-year risk of the outcome under treatment and no treatment (and maintain the intended treatment effects), we used a root-finding algorithm to solve for the discrete time outcome probabilities (Methods S2).

Based on the two visits over which individuals could contribute indexes, there were three potential treatment combinations: no treatment at Visit 1 or Visit 2, no treatment at Visit 1 and treated at Visit 2, and treated at Visit 1 and Visit 2. To allow full follow-up for indexes contributed at Visit 2, we derived each simulated individual's three-year potential outcomes and times-to-event under each of these possible treatment patterns (Methods S3, Figure S2).

*2.4.2. Simulation Samples and Treatment Assignment*

For each of the 5,000 simulations in each scenario, we randomly sampled 5,000 individuals with replacement from the superpopulation.

Once each simulated cohort was selected, we generated OCA treatment values for the SPT and the emulated SNTs. OCA treatment for the SPT was generated marginally at Visit 1. OCA treatment for the emulated SNTs was generated at Visit 1, conditional on PBC severity. OCA treatment at Visit 2 was generated conditional on PBC severity and whether Visit 2 qualified as a treatment decision point.

If an individual in the emulated SNT was treated with OCA at Visit 1, OCA treatment was carried forward to Visit 2. All treatment values at Visit 2 were carried forward to Visit 3. Observed outcomes for the SPT and the emulated SNTs were assigned based on the potential outcomes derived in the superpopulation corresponding to the individual's observed treatment pattern (Methods S3).

*2.4.3. Selection and Structure of Data*

*"Truth".* The true treatment effect for each scenario was calculated as the contrast between the potential outcomes across the entire superpopulation under no treatment at Visits 1 and 2 and treatment at Visits 1 and 2. This contrast corresponds to the effect had everyone received treatment at initial eligibility (Visit 1) versus had no one received treatment through Visit 2.

*Single Point Trial (SPT).* All simulated individuals contributed one index at Visit 1. The observed outcome was the potential outcome corresponding to their assigned treatment at Visit 1.

*Calendar SNT Emulation (eSNT-CAL).* All simulated persons were indexed into the study at Visit 1. OCA non-initiators contributed indexes at Visit 2. If treated at Visit 2, non-treated Visit 1 indexes were censored at Visit 2. The time to outcome from Visit 2 was calculated as the time between Visit 2 and the time of the observed outcome (Figure S1, Methods S3).

*Treatment Decision Point SNT Emulation (eSNT-TD).* All simulated persons were indexed into the study at Visit 1. OCA non-initiators contributed indexes at Visit 2 if it qualified as a treatment decision point. Censoring and outcome ascertainment for each contributed index followed the same derivation as the eSNT-CAL.

### 2.5. Scenarios

In the eSNT-CAL and eSNT-TD, the probability of treatment at each Visit was 0.25 and 0.75 for low and high PBC severity, respectively. In the SPT, the probability of treatment at Visit 1 was equal to the marginal probability of treatment for the eSNT-CAL/eSNT-TD (~37.5%).

We varied two data generation parameters across scenarios: the effect of disease severity on the probability of a treatment decision point (null and non-null) and the treatment effect (modification by disease severity and no modification on the ratio scale) (Figure 2), resulting in four scenarios. Plainly, the four scenarios are:
    (1) No effect of PBC severity on meeting treatment decision criteria, homogeneous treatment effect;
    (2) No effect of PBC severity on meeting treatment decision criteria, heterogeneous treatment effect;
    (3) Effect of PBC severity on meeting treatment decision criteria, homogeneous treatment effect; and
    (4) Effect of PBC severity on meeting treatment decision criteria, heterogeneous treatment effect
For each scenario, we sampled 5,000 cohorts of 5,000 individuals from the superpopulation.

*2.5.1. Scenarios by the Effect of Severity on Treatment Decision Points*

By design, all simulated patients contributed an index at Visit 1. For scenarios where PBC severity did not affect the occurrence of a treatment decision point, we specified the probability of a treatment decision point at Visit 2 as 0.3. For scenarios where PBC severity did affect the probability of a treatment decision point, the probabilities were 0.2 and 0.8 for low and high severity, respectively.

*2.5.2 Scenarios by Treatment Effect*

Regardless of scenario, the two-year outcome probability among OCA non-initiators was 15% for low severity and 25% for high severity visits. When the treatment effect was not modified by PBC severity, the two-year risk ratio (RR) for the composite outcome comparing treatment to no treatment was 0.7. When modified by severity, the two-year RRs were 0.5 and 0.9 for low and high severity, respectively.

### 2.6. Analyses

We described the simulated study participants and their indexes for each scenario and study design by calculating the median and interquartile range for each summary statistic across all 5,000 cohorts.

In the SPT, we estimated the crude RR using simple proportions of the outcome among those in each treatment group at baseline. Due to randomization, crude effect measures are expected to be unbiased estimates of the ATE and ATT. However, to facilitate comparison with the standardized estimates from the SNT study designs, we directly standardized the crude risks to the distribution of disease severity among all and treated trial participants to target the ATE and ATT, respectively.[22]

The eSNT-CAL and eSNT-TD simulated data were subject to confounding by disease severity and selection bias due to artifical right censoring of non-OCA indexes due to OCA initiation at Visit 2. To address informative right censoring, we estimated risks via inverse probability of censoring weighted (IPCW)[23] Kaplan-Meier estimators (Methods S1). To address confounding, we used non-parametric direct standardization on baseline disease severity in the target population.[22] Estimated risks were contrasted via the RR.

We first estimated the "crude" RR estimates corrected only for selection bias arising from informative right censoring. Next, we standardized the crude RR estimates to: (1) the severity distribution among *all* indexes (i.e., eSNT-CAL/eSNT-TD ATE) and (2) the severity distribution among *treated* indexes (i.e., eSNT-CAL/eSNT-TD ATT; Table S1). Finally, we standardized the RRs to (1) the severity distribution among all SPT participants (i.e., SPT ATE) and (2) severity distribution among SPT participants randomized to treatment (i.e., SPT ATT).

We then summarized the bias, root mean-squared error, and empirical standard error for the crude and standardized log-RRs using the formulae in Table S2.[25] Here, bias refers to discrepancies of the estimates from each study design compared to the SPT "truth". Specifically, we define bias as:

$$\widehat{Bias}(\hat{\theta}_{a,e}) = E(\hat{\theta}_{a,e}) - \theta_{SPT,e} = \frac{1}{5000} \sum_{i=1}^{5000} \hat{\theta}_{a,e,i} - \theta_{SPT,e}$$

where $a \in \{SPT, eSNT\text{-}CAL, eSNT\text{-}TD\}$ and $e \in \{ATE, ATT\}$ together denotes specific estimand being targeted by the estimated treatment effect $\hat{\theta}_{a,e}$. To calculate bias, we always compared to the known "truth" from the single point trial, $\theta_{SPT,e}$, which we calculated from the potential outcomes in the super population.

All analyses were conducted in R version 4.5.0, using tidyverse packages.[26] Code is available on GitHub (https://anonymous.4open.science/r/publications-00BF/).

## 3. Results

The distribution of high (25%) versus low (75%) PBC severity was the same in the SPT for both the ATE and the ATT across all scenarios. Because PBC severity could only worsen over time, the implied target populations of the collection of all indexes in the eSNT-CAL and eSNT-TD had a higher proportion of high severity PBC compared to the SPT (Table 2). When severity impacted the probability of a treatment decision point (Scenarios 3 and 4), the treated indexes from the SNT-TD (targeting the ATT in the SNT-TD) represented the most discrepant target population from the SPT, with 70% of indexes contributed by individuals with high severity PBC.

When the treatment effect was homogeneous across levels of PBC severity, addressing confounding bias removed the difference between the eSNT-CAL/TD estimates and the reference "truth", resulting in unbiased estimates.

When PBC severity modified the treatment effect (Scenarios 2 and 4) discrepancies remained in the RR estimates even after removing confounding bias in the eSNT-CAL/TD estimates, compared to the RR in the SPT. Across scenarios, the treated eSNT-CAL/TD indexes had higher severity than all eSNT-CAL/TD indexes, which caused estimates targeting the ATT to be more biased than those targeting the ATE (log-RR bias range: 0.04-0.09 for the eSNT-CAL/TD ATE versus 0.14-0.17 for the eSNT-CAL/TD ATT) (Table S4).

Standardizing risk estimates from the eSNT-CAL/TD to the PBC severity distribution from the SPT (at Visit 1) effectively removed bias in scenarios with effect measure modification by severity (Scenarios 2 and 4) (Figure 3).

The empirical standard error (ESE) ranged from 0.51 to 0.53 for the SPTs (Table S4). The ESE was lowest for eSNT-CAL estimates, crude or standardized to the PBC severity distribution in eSNT-CAL population (range: 0.40 to 0.48); this range was higher for the eSNT-TD estimates (range: 0.44 to 0.57). Once standardized to the PBC severity distribution in the SPT population, ESE for the eSNT-CAL estimates increased to about the range for the SPT estimates (range: 0.49 to 0.56). We observed a slightly elevated ESE range among estimates from the eSNT-TD standardized to the PBC distribution in the SPT (range: 0.52 to 0.58).

Root mean squared error (rMSE) was the highest among the confounding-corrected estimates most discrepant from the SPT contrast, ranging from 0.15-0.19 in the SNT emulations targeting the SNT ATT in scenarios with effect measure modification. The rMSE was more modest in the SNT ATE, but greatest (0.11) in the SNT emulations with effect measure modification and differential treatment decision points by severity (Scenario 4). Standardizing the eSNT estimates to the first time point in the SPT removed the bias and increased ESE, resulting in rMSE approximately equal to the rMSE of the SPT (0.07-0.08).

## 4. Discussion

This study investigated the implicit target population of sequential nested trial emulations and how it may result in different treatment effect estimates than what we would derive from a conventional single point trial. In settings without effect measure modification by disease severity, ATE and ATT RR estimates in all studies were unbiased after correcting for confounding. However, when effect measure modification was present, confounding-adjusted RR estimates from the eSNT-CAL and eSNT-TD remained discrepant from the SPT. Findings were consistent regardless of whether we incorporated treatment decision points into inclusion criteria. Finally, while the eSNT-CAL design resulted in the lowest ESE after adjusting for confounding, ESEs were comparable to the other study designs once results were standardized to the population from the SPT.

We considered all (ATE) and treated (ATT) indexes as target populations for our SNT-CAL/TD analyses to explore their comparative performance. In estimates targeting the ATE, the implied population is the collection of indexes contributed by all individuals and reflects a changing distribution of disease severity over time. This causal estimand is appealing because it is the target estimand of a single point randomized trial. However, the target population from the SNT is unclear compared to the single point trial: the analytic units are indexes, not people. Estimates targeting the ATT, a population where each individual contributes only one index at the time they receive treatment, might be an easier population to describe. However, this study illustrated that both populations can return biased results when compared to the single point trial. In fact, in some cases, estimates from the ATT may be even more biased than those from the ATE.

Sequential nested trial emulations were introduced as an efficient way to use longitudinal observational data and avoid immortal time bias by allowing individuals to contribute multiple time zeroes to analyses.[17,28] However, early adopters did not specify the target population of the effects estimated from the study design.[29] With the advent of the target trial framework,[3] more recent applications of the sequential nested trial emulation define a causal estimand in the context of a randomized controlled trial.[4,12,13,30] Even so, the target population typically implied by the analysis is the collection of all contributed indexes, which is not reflective of a real-world population that exists at any one point in time. In our simulation, we considered the single point trial as our "truth" because the implied target population is clear: everyone randomized at baseline. Although we indexed the same population at Visit 1 in all designs (SPT, eSNT-CAL, and eSNT-TD), sequential indexing in the eSNT-CAL and eSNT-TD resulted in shifts in the implied target population. As a result, the confounding-adjusted RRs from these designs deviated from those of the SPT and from each other. Given that randomized single point trials have been the hallmark of causal inference and clinical and public health decision making, interpreting estimates from sequential nested trial emulations (with or without the treatment decision design) may not be immediately clear for stakeholders.

Our study has notable strengths. Foremost, it builds on prior work aiming to improve the utility of sequential nested trial emulation for causal analyses using observational data. Keogh et al. showed that estimates from a sequential nested trial emulation and inverse probability weighted marginal structural models can target the same causal estimand when the target population is clearly defined; in simulations and in their applied example, they standardized pooled results to the first time point contributed to the sequential trials.[29] While their focus was on per-protocol effects under sustained treatment, our study extends this framework by comparing ATE and ATT across study designs and illustrating the consequences of effect modification when populations differ. Further, we grounded our data generating mechanisms in patterns we might expect in real data and compared our results to the trial that we typically want to conduct in the real world: a single point trial. Ultimately, our results reinforce the importance of clearly specifying both the causal estimand and the target population for interpretation.

However, like all statistical simulations, our data are a simplification of reality. Sequential nested trial emulation with reweighting to the corresponding single point randomized trial requires that investigators account for effect measure modifiers. Because we used simulated data, these variables were perfectly captured; however, we would expect at least some of such variables to be unmeasured or imperfectly measured in real-world data sources,[31] resulting in lingering bias.[32–34] Further, using the sequential nested trial design to target the implied SPT estimand requires that the time from the initial index does not modify the treatment effect; if such modification were present, including any indices after the initial index would bias the results. Finally, our "homogeneous treatment effect" scenarios were only homogeneous on the risk ratio scale: investigators should take into account the scale of their measure when assessing for effect measure modifiers.

Our study additionally expands upon prior work by considering sequential nested trial emulation with a treatment decision design.[18] It poses many of the same strengths as the calendar-based SNT for estimating the comparative effect of two treatment regimens. First, it maximizes investigators' ability to capture treatment effects by retaining all index events. While this may pose computational challenges when individuals contribute a large number of untreated indexes,[3,4,16] this fact is important when treatment is rare.[35] Second, it facilitates avoiding immortal time bias in studies with a non-active comparator.[3,4,16] However, an additional strength of the treatment decision design over the calendar-based approach is that it allows investigators to define a target population that may more clearly align with clinical decision-making. This study illustrated that both approaches are valid so long as the target populations are carefully considered.

## 5. Conclusion

Studies using sequential nested trial emulation should carefully consider how the distribution of treatment-modifying comorbidities change over time and, subsequently, over indexes within a person. In fact, changes in the distribution of effect measure modifiers should be expected, as higher risk people are likely to have events early, so later indexes may be more heavily skewed towards lower risk people. Such changes could result in estimates that meaningfully differ from those that we would expect from the trial we would do in practice: a single point trial. While expected, this reality complicates interpretation of the effect estimates. It is unlikely we would conduct the target trial of sequential nested trial emulation in practice, making the target population and applied intervention abstract and poorly defined. When reporting descriptive statistics of the study population, this should likely be done on the index-level both for all contributed indexes and those relevant to the target estimands so that readers can evaluate the distribution of potential effect measure modifiers in the analyses. Further, investigators should consider whether it is appropriate to reweight their results to a clearly defined target population, such as the single point randomized trial. Ultimately, investigators must consider the target population implied by their study and clearly articulate that for adequate generalization of findings to real-world populations.

**Tables and Figures**

**Table 1.** Protocol components of the target trials for the single point trial and the calendar-based sequential nested trial (SNT-CAL) and the treatment decision-based sequential nested trial (SNT-TD).

| Component | Single Point Trial | SNT-CAL | SNT-TD |
|---|---|---|---|
| **Aim** | To estimate the effect of initiating versus not initiating obeticholic acid (OCA) on risk of composite death, liver transplant, or hospitalization for hepatic decompensation among patients with primary biliary cholangitis (PBC). | Same. | Same. |
| **Eligibility** | | Same as SPT, plus: | Same as SPT, plus: |
| **Criteria 1** | • Diagnosed PBC and not currently using OCA | • Patients are eligible to be re-randomized one-year after initial index if they did not receive OCA at Visit 1. | • Patients are eligible to be re-randomized one-year after initial index if they did not receive OCA at Visit 1 and meet the treatment decision criteria. |
| **Criteria 2** | • Diagnosed PBC and not currently using OCA and<br>• Laboratory values indicating a need for pharmacotherapy | | |
| **Treatment strategies** | 1. Initiate and continue treatment with OCA.<br>2. Never initiate OCA. | Same. | Same. |
| **Treatment assignment** | Random assignment to one of the treatment strategies.<br><br>Patients will be aware of their assigned treatment. | Same. | Same. |
| **Follow-up** | Follow-up will begin at randomization and continue until the first of the study outcome, deviation from assigned treatment strategy, or the end of follow-up (two years). | Same. | Same. |
| **Outcome** | The first of either death, liver transplant, or hospitalization due to hepatic decompensation. | Same. | Same. |
| **Causal contrast** | Per-protocol effect: Effect of initiating versus never initiating OCA. | Same. | Same. |
| **Statistical Analysis** | Per-protocol analysis. | Same. | Same. |

**Table 2.** Target population implied by the distribution of severity of the indexes included in the single point trial, sequential nested trial emulation (eSNT-CAL), and the sequential nested trial emulation with a treatment decision design (eSNT-TD) averaged across the 5,000 simulated cohorts.

|  | Single point trial | | eSNT-CAL | | eSNT-TD | |
| --- | --- | --- | --- | --- | --- | --- |
|  | All indexes | Treated indexes | All indexes | Treated indexes | All indexes | Treated indexes |
| *Scenario 1: severity does not impact visit cadence; no effect modification* | | | | | | |
| Low severity, % | 75 | 75 | 56 | 40 | 67 | 40 |
| High severity, % | 25 | 25 | 44 | 60 | 33 | 60 |
| *Scenario 2: severity does not impact visit cadence; effect modification by severity* | | | | | | |
| Low severity, % | 75 | 75 | 56 | 40 | 67 | 40 |
| High severity, % | 25 | 25 | 44 | 60 | 33 | 60 |
| *Scenario 3: severity impacts visit cadence; no effect modification* | | | | | | |
| Low severity, % | 75 | 75 | 56 | 30 | 56 | 30 |
| High severity, % | 25 | 25 | 44 | 70 | 44 | 70 |
| *Scenario 4: severity impacts visit cadence; effect modification by severity* | | | | | | |
| Low severity, % | 75 | 75 | 56 | 30 | 56 | 30 |
| High severity, % | 25 | 25 | 44 | 70 | 44 | 70 |

**Figure 1.** Directed acyclic graphs used to generate the underlying patient data for (A) the single point trials and (B) the sequential nested trial emulations. Variables are indexed by the year *t* at which the value was observed. The node "Treatment Decision" represents the simulated individual having an encounter with a sufficiently high laboratory value to indicate treatment with obeticholic acid (OCA). The dashed arrow between "Eligible Encounter$_t$" and "Treatment$_t$" represents a deterministic relationship where treatment with OCA was only possible if the simulated individual experienced a treatment decision point.

**(A) Single point trial**

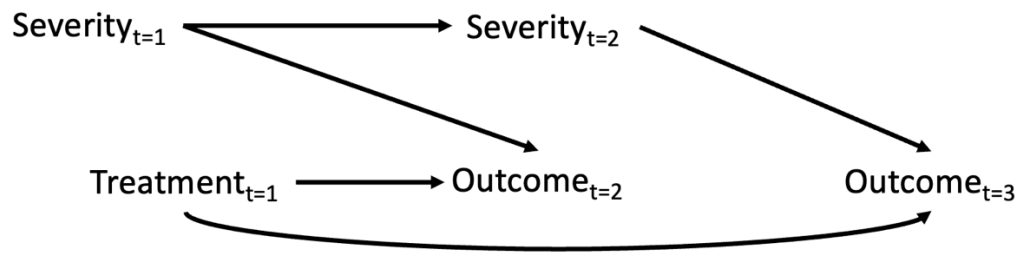

**(B) Sequential nested trial emulation**

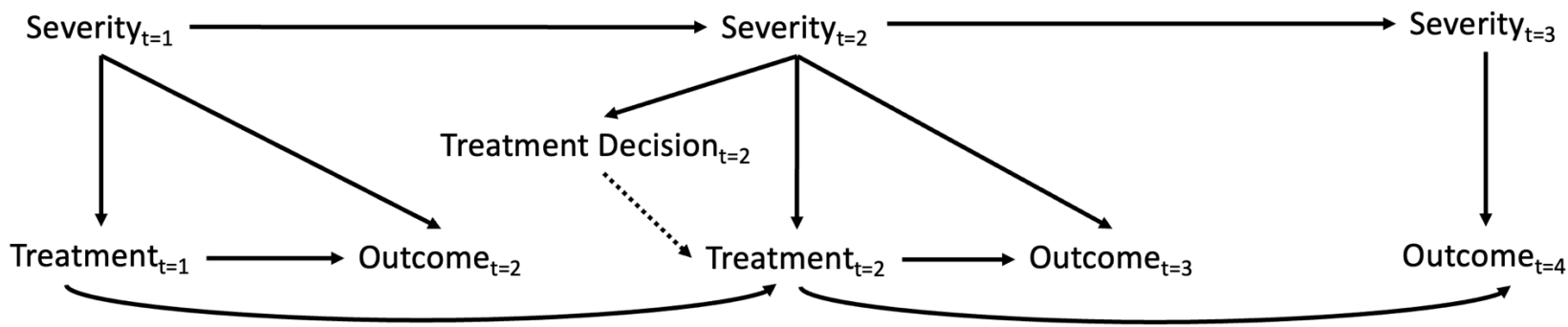

**Figure 2.** Schematic of the scenarios generated in this study.

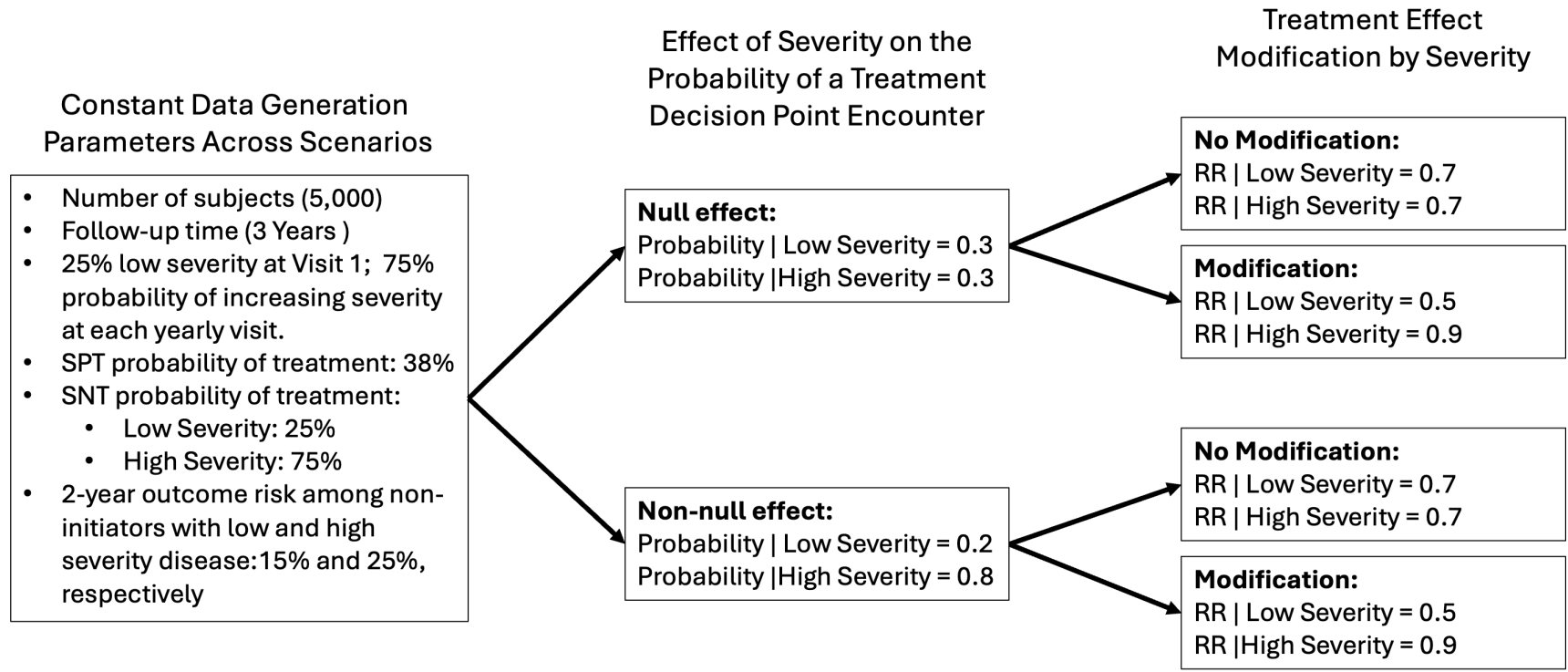

**Figure 3.** Bias in the log-transformed risk ratio estimates for the average treatment effect among the entire population (ATE) from the single point trial (SPT), calendar-based sequential nested trial emulation (eSNT-Cal), and the SNT emulation with a treatment decision design (eSNT-TD). Scenarios are distinguished by disease severity having a null (1, 2) and non-null (3, 4) effect on the probability of a treatment decision point and by disease severity not modifying (1, 3) and modifying (2, 4) the treatment effect. The shape indicates the target population to which the risk estimates were standardized for confounding adjustment wherein "crude" implies no non-parametric standardization, "SNT" implies the indexes of the SNT, and "SPT" implies all participants in the single point trial.

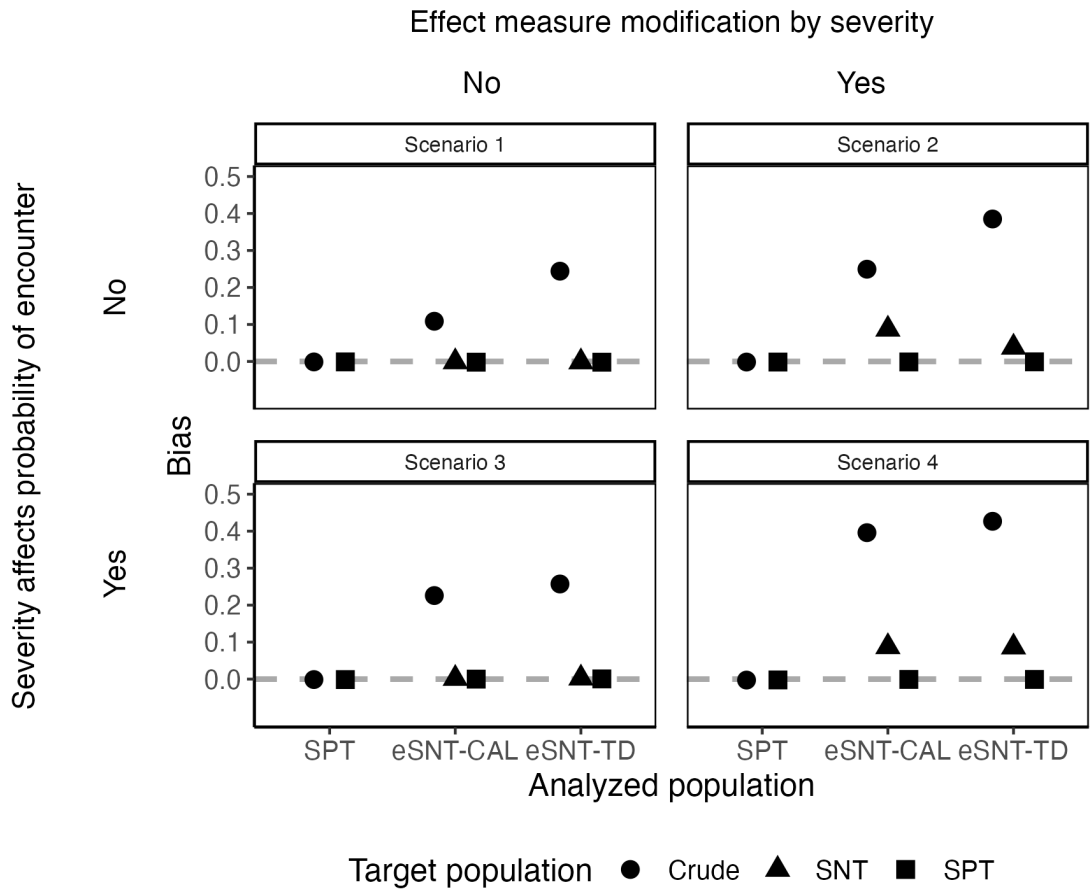

**Supplemental Document**

# Table of Contents

## Methods

**Methods S1.** Deriving inverse probability of artificial censoring weights.

Artificial censoring of untreated indexes at treatment initiation was informative due to time-varying severity: disease severity drove both treatment decision points and treatment initiation. We constructed inverse probability of artificial censoring weights from the known conditional probabilities of treatment decision points and treatment initiation.

We generated study data over three annual visits. Individuals could contribute indexes over the first two visits. By design, all individuals contributed an index at Visit 1. Therefore, untreated indexes had at most one additional time point at which they could be treated and artificially censored. Specifically, untreated indexes from Visit 1 could initiate treatment at Visit 2. Individuals could not initiate treatment at Visit 3.

We constructed inverse probability of artificial censoring weights using the steps outlined below. We describe these steps in detail for the sequential nested trial (SNT) emulation with a treatment decision design and then indicate how this would be simplified for the calendar-based SNT. Let $k = 1, 2, 3$ indicate visit, $L_k = 0,1$ indicate low versus high severity primary biliary cholangitis (PBC) at Visit $k$, $A_k = 0,1$ indicate not treated versus treated at Visit $k$, and $V_k = 0,1$ indicate whether an individual did or did not meet treatment eligibility (i.e., had a treatment decision point) at Visit $k$.

Let $t = 1, 2$ denote one-year chunks of follow-up after each contributed index. We aimed to estimate risk at $\tau = 2$ (2-years). Let $C_t = 0, 1$, where 0 indicates that an index is uncensored at $t$.

By design, treated indexes could not be artificially censored from treatment. Therefore, their inverse probability of censoring weights were equal to 1 across follow-up.

For untreated indexes contributed at Visit $k$, we defined the probability of remaining untreated at each follow-up Visit as $\Pr(A_{k+t-1} = 0|L_{k+t-1})$ and the probability of having an eligible index as $\Pr(V_{k+t-1} = 1|L_{k+t-1})$. As treatment and treatment eligibility were generated independently, the conditional probability of remaining untreated, and thus uncensored, at each Visit is equal to:

$$\Pr(C_t = 0|L_{k+t-1}) = 1 - (\Pr(A_{k+t-1} = 1|L_{k+t-1}) \times \Pr(V_{k+t-1} = 1|L_{k+t-1}))$$

For the calendar-based SNT emulation, this probability would be calculated by removing the $\Pr(V_{k+t-1} = 1|L_{k+t-1})$ term (i.e., $1 - (\Pr(A_{k+t-1} = 1|L_{k+t-1}))$).

The inverse probability of artificial censoring weights at each 1-year of relative follow-up $t$ were then calculated as the inverse of the cumulative probability of remaining uncensored up to that time point:

$$IPCW_t = \frac{1}{\prod_{j=1}^{t} \Pr(C_t = 0|L_{k+t-1})}$$

**Methods S2.** Derivation of discrete conditional composite outcome probabilities

The binary composite outcome could occur either at one or two years post-index. To maintain our *a priori* defined two-year cumulative risk of the outcome under each treatment and severity level, we derived discrete time specific outcome probabilities conditional on severity and treatment. To do this, we solved a system of four equations for the four discrete time unknown probabilities: the probability of the outcome under no treatment and low severity; the probability under treatment and low severity; the probability under no treatment and high severity; and the probability under treatment and high severity.

- Let $p_{a,z}$ denote the discrete time risk of the outcome under treatment level $a \in (0,1)$ and PBC severity $z \in (0,1)$.
- Let $\pi$ denote the probability of PBC severity increasing to high severity.
- Let $F_{Y|a(1),z(1)}(2)$ denote the two-year risk of the outcome, where $a(1)$ and $z(1)$ represent the treatment and PBC severity levels at the index Visit.
- Let $\delta_{z(1)}$ denote the two-year treatment effect for PBC severity $z(1)$.

In words, the two-year risk of the outcome is equal to the probability of the outcome one-year after index, plus the probability of the outcome at year two after index, given that the outcome didn't occur at year one. For the two-year risks when the initial index is at low severity, we have to account for the probability of increasing severity one year after index to appropriately estimate the discrete time probabilities.

This corresponds to the following four equations:
1. Two-year risk of the outcome, under high severity and no treatment:
$$F_{Y|0,1}(2) = p_{0,1} + (1 - p_{0,1})p_{0,1}$$
2. Two-year risk of the outcome, under high severity and treatment
$$F_{Y|1,1}(2) = \delta_1^{-1}(p_{1,1} + (1 - p_{1,1})p_{1,1})$$
3. Two-year risk of the outcome, under low severity and no treatment
$$F_{Y|0,0}(2) = p_{0,0} + (1 - \pi)(1 - p_{0,0})p_{0,0} + \pi(1 - p_{0,0})p_{0,1}$$
4. Two-year risk of the outcome, under low severity and treatment
$$F_{Y|1,0}(2) = \delta_0^{-1}(p_{1,0} + (1 - \pi)(1 - p_{1,0})p_{1,0} + \pi(1 - p_{1,0})p_{1,1})$$

We rearranged and set each of the four equations to 0. Substituting the scenario agnostic cumulative risks and increased severity probability, as well as the scenario specific treatment effects, we implemented a root-finder to solve the system of equations for the four discrete time probabilities. Specifically, we used the Newton-Raphson root-finding algorithm in R (using the rootSolve::multiroot function).

**Methods S3.** Treatment patterns and potential outcomes.

Based on two Visits where individuals could receive treatment, two treatment levels (initiated versus not initiated), and assumed fidelity to treatment once initiated, each individual had three possible treatment patterns:

|  | Treated with OCA at Visit 1? | Treated with OCA at Visit 2? |
|---|---|---|
| $\bar{a}_1 = (0, 0)$ | No | No |
| $\bar{a}_2 = (0, 1)$ | No | Yes |
| $\bar{a}_3 = (1, 1)$ | Yes | Yes |

As part of the initial data generation for the superpopulation, we generated potential outcomes at each time point under each treatment level $a \in (0,1)$, conditional on an individual's PBC disease severity $Z$ at that time point.

- Let $Y^1(t + 1)|Z(t)$ be the potential outcome at time $t + 1$ under treatment at time $t$, conditional on Z at time $t$;
- Let $Y^0(t + 1)|Z(t)$ be the potential outcome at time $t + 1$ under no treatment at time $t$, conditional on Z at time $t$

Then, under each possible treatment pattern, we derived the *Visit* the first outcome occurred. For example, considering the following individual's generated potential outcomes:

|  | $Y^0(Visit)$ | $Y^1(Visit)$ |
|---|---|---|
| **Visit 2** | 0 | 1 |
| **Visit 3** | 1 | 0 |

We can derive the corresponding potential times to outcomes $T^{\bar{a}}$ under each treatment pattern:

|  | **Outcome Visit** | $T^{\bar{a}}$ |
|---|---|---|
| $\bar{a}_1 = (0, 0)$ | Visit 3 | 2 years (Visit 3 – Visit 1) |
| $\bar{a}_2 = (0, 1)$ | No outcome | - |
| $\bar{a}_3 = (1, 1)$ | Visit 2 | 1 year (Visit 2 – Visit 1) |

Then, after we generated treatment values at Visits 1 and 2, we assigned the potential "time to outcome" compatible with their treatment pattern for the observed outcome.

**Figures**

**Figure S1.** Depiction of the index events included in the single point trial versus the sequential nested trial.

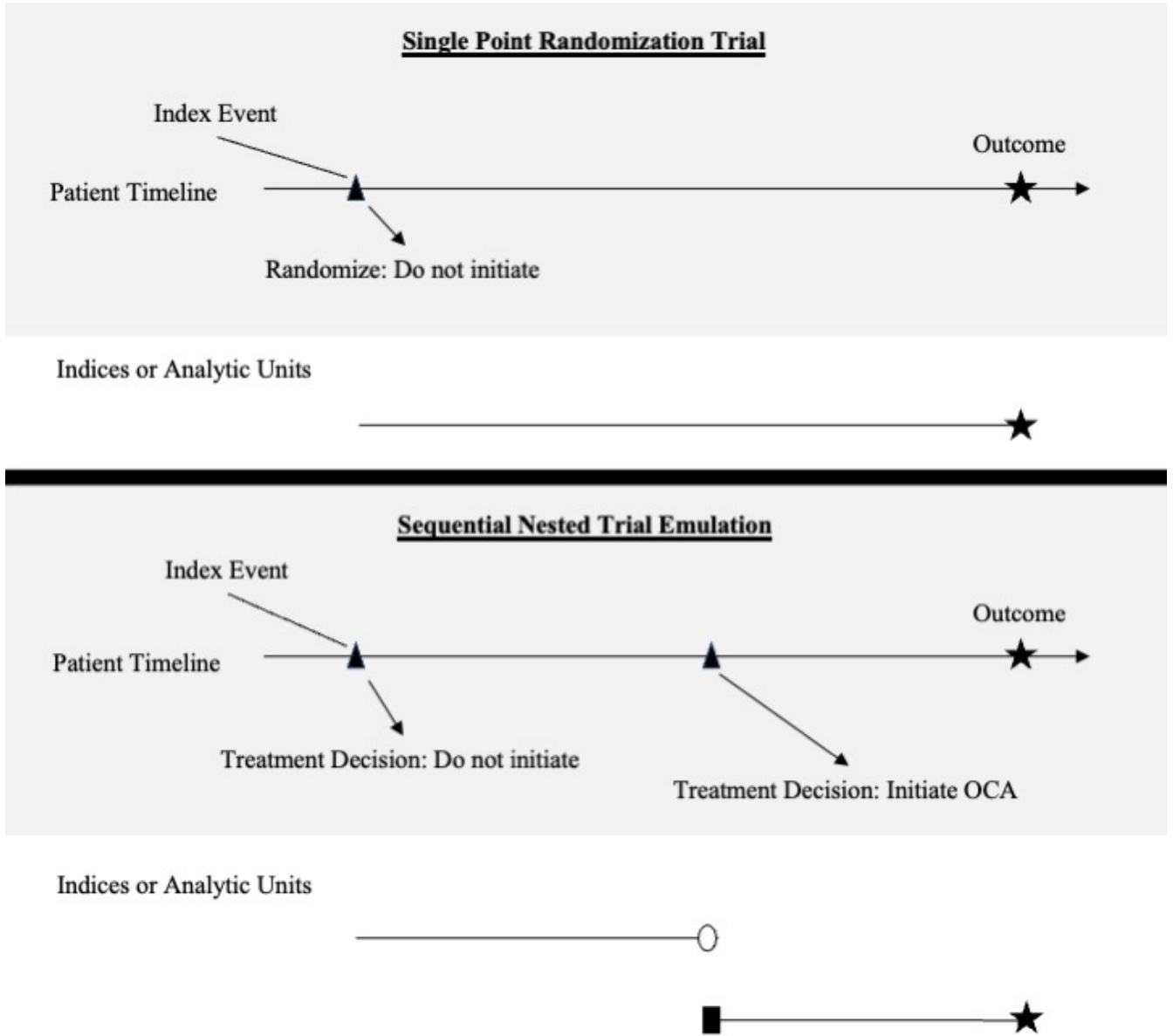

**Figure S2.** Depiction of the data generation process for each of the study designs. We generated data at 3 discrete visits: the initial index (Visit 1) and then two subsequent visits (Visits 2-3). Disease severity, designation as a treatment decision point, and treatment were observed at that visit. Outcomes were determined at that visit but observed 1 year later (e.g., an outcome of 1 at Visit 1 implied that the simulated patient experienced the outcome at Visit 2).

| | Visit 1 | Visit 2 | Visit 3 |
|---|---|---|---|
| **PBC severity** | 0 | 0 | 1 |
| **Treatment decision?** | - | 0 | - |
| **Visit-specific potential outcome under non-initiation** | 1 | 0 | 1 |
| **Visit-specific potential outcome under initiation** | 0 | 1 | 1 |
| **Potential outcome under initiation at Visit 1** | 0 | 1 | - |
| **Potential outcome under initiation at Visit 2** | 1 | - | - |
| **Potential outcome under never initiate** | 1 | - | - |
| **Treatment** | 0 | 1 | - |
| **Observed outcome** | 1 | - | - |

**Step 1.** Generate time-varying PBC severity (Visits 1-3)

**Step 2.** Generate whether Visit 2 meets criteria for a treatment decision

**Step 3.** Generate visit-specific potential outcomes (Visits 1–3)

**Step 4.** Determine potential outcome under each treatment pattern

**Step 5.** Generate treatment values at Visit 1 and Visit 2

**Step 6.** Determined realized outcome under actual treatment value.

**Figure S3.** Bias in the log-transformed risk ratio estimates for the average treatment effect among the treated (ATT) from the single point trial (SPT), calendar-based sequential nested trial emulation (eSNT-Cal), and the SNT emulation with a treatment decision design (eSNT-TD). Scenarios are distinguished by disease severity having a null (1, 2) and non-null (3, 4) effect on the probability of a treatment decision point and by disease severity not modifying (1, 3) and modifying (2, 4) the treatment effect. The shape indicates the target population to which the risk estimates were standardized for confounding adjustment wherein "crude" implies no non-parametric standardization, "SNT" implies the indexes of the SNT, and "SPT" implies all participants in the single point trial.

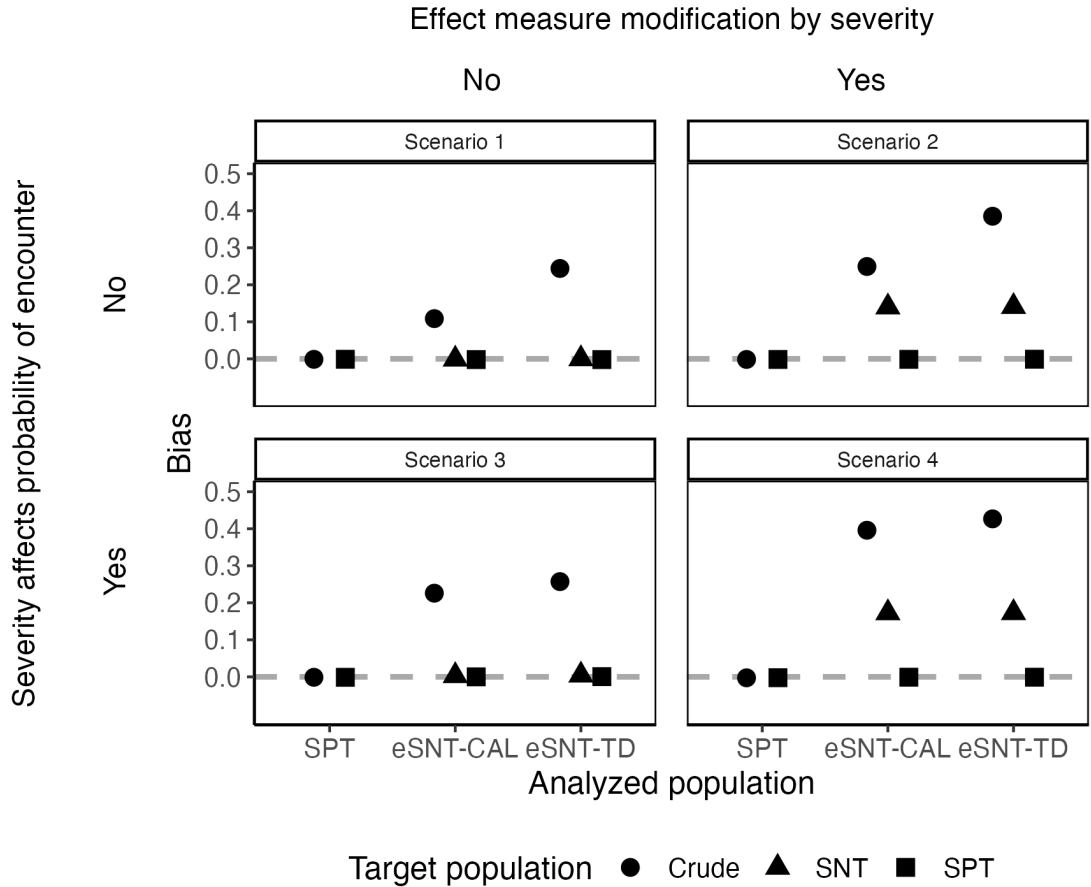

# Tables

**Table S1.** Tabular description of the analyses conducted within each cohort and study design to estimate the effect of obeticholic acid (OCA) initiation (versus non-initiation) on the risk of the composite study outcome.

| Study Design | Analysis | Description |
| --- | --- | --- |
| Single Point Trial | True risk ratio (RR) | We calculated the OCA-initiation risk by dividing the number of events that would have been observed if all indexes had initiated OCA by the number of indexes included in the trial. We calculated the non-initiation risk similarly considering the outcomes that would have occurred if everyone had not initiated OCA. We then contrasted these risks via the risk ratio. |
|  | Crude RR | Using simple proportions, we calculated the risk of the study outcome among the treated indexes and the untreated indexes, separately. We then contrasted these risks by dividing the risk among the treated by the risk among the untreated indexes. |
|  | RR standardized to all indexes (i.e., ATE) | We used non-parametric direct standardization to calculate a risk ratio standardized to the disease severity distribution (measured at the index) among all indexes.[1]<br><br>Specifically, we calculated the risk of the outcome within each stratum of primary biliary cholangitis (PBC) severity at the index and then weighted these risks according to the distribution of PBC severity in the total population (i.e., all indexes included in the trial). |
|  | RR standardized to the treated indexes (i.e., ATT) | We used non-parametric direct standardization to calculate a risk ratio standardized to the PBC severity distribution (measured at the index) among treated indexes. |

|  |  | Specifically, we calculated the risk of the outcome within each stratum of PBC severity at the index and then weighted these risks according to the distribution of PBC severity in the treated population (i.e., all treated indexes included in the trial). |
| --- | --- | --- |
| SNT Emulations (Regardless of Inclusion Criterion for Treatment Decision Points) | Inverse probability of censoring weighted RR (i.e., "crude" RR) | We estimated the risk of the study outcome in each treatment arm using a Kaplan-Meier estimator with inverse probability of censoring weights (IPCW). IPCW are time-varying weights, updated at each week of follow-up, that weight uncensored individuals according to their probability of not being censored so that they can stand in for censored individuals. Because censoring model misspecification was not of-interest in this study, we used the known values for these probabilities. |
|  | RR standardized to all indexes in the SNT emulation (i.e., ATE in the SNT emulation) | We used a Kaplan-Meier estimator with IPCW to calculate the IPC weighted 1-year risk among indexes within each strata of PBC severity (at the index) and treatment. We then weighted these risks according to the PBC severity distribution of all indexes included in the SNT emulation using non-parametric direct standardization. |
|  | RR standardized to the treated indexes in the SNT (i.e., ATT in the SNT) | Risk estimation was the same as above. We then weighted these risks according to the PBC severity distribution of all treated indexes included in the SNT emulation using non-parametric direct standardization. |
|  | RR standardized to all indexes in the single point trial (i.e., ATE in the single point trial) | Risk estimation was the same as above. We then weighted these risks according to the PBC severity distribution of all indexes included in the single point trial using non-parametric direct standardization. |

| | RR standardized to treated indicated in the single point trial (i.e., ATT in the single point trial) | Risk estimation was the same as above. We then weighted these risks according to the PBC severity distribution of treated indexes included in the single point trial using non-parametric direct standardization. |

ATE = Population average treatment effect; ATT = Average treatment effect among the treated.

**Table S2.** Performance measures estimated for our simulation using formulae from Morris et al. 2019.[2]

| Performance Measure | Definition | Numeric Definition[a] | Estimate | Monte Carlo Standard Error |
|---|---|---|---|---|
| **Bias** | The amount by which the estimated effect measure differs from the true effect measure | $E[\hat{\theta}] - \theta$ | $\frac{1}{n_{sim}} \sum_{i=1}^{n_{sim}} \hat{\theta}_i - \theta$ | $\sqrt{\frac{1}{n_{sim}(n_{sim}-1)} \sum_{i=1}^{n_{sim}} (\hat{\theta}_i - \bar{\theta})^2}$ |
| **Empirical standard error** | Measures the precision of estimated effect measure | $\sqrt{Var(\hat{\theta})}$ | $\sqrt{\frac{1}{n_{sim}-1} \sum_{i=1}^{n_{sim}} (\hat{\theta}_i - \bar{\theta})^2}$ | Did not calculate. |
| **Mean-squared error** | Sum of squared bias and variance of the estimated effect measure | $E[(\hat{\theta} - \theta)^2]$ | $\frac{1}{n_{sim}} \sum_{i=1}^{n_{sim}} (\hat{\theta}_i - \theta)^2$ | Did not calculate. |

[a] $\theta$ refers to the true value of the treatment effect parameter (in this study, the true risk ratio).

**Table S3.** Descriptive statistics on the simulated patients and indexes included in the single point trial, sequential nested trial emulation (eSNT-CAL), and the sequential nested trial emulation with a treatment decision design (eSNT-TD) across the 5,000 simulated cohorts.

| Summary measure | Single point trial Median (IQR) Non-initiators | Initiators | eSNT-CAL Median (IQR) Non-initiators | Initiators | eSNT-TD Median (IQR) Non-initiators | Initiators |
|---|---|---|---|---|---|---|
| *Scenario 1: severity does not impact visit cadence; no effect modification* | | | | | | |
| Number of people | 3120 (3096, 3144) | 1880 (1856, 1904) | 2520 (2498, 2547) | 2480 (2453, 2502) | 2520 (2498, 2547) | 2480 (2453, 2502) |
| Number of indexes | | | | | | |
|     Low severity | 2335 (2313, 2365) | 1413 (1389, 1429) | 2900 (2866, 2930) | 1521 (1494, 1543) | 2434 (2411, 2459) | 1521 (1494, 1543) |
|     High severity | 788 (767, 801) | 467 (451, 483) | 2018 (1979, 2042) | 1528 (1502, 1548) | 412 (400, 423) | 1528 (1502, 1548) |
| Percent high severity indexes | 25 (25, 26) | 25 (24, 26) | 41 (41, 41) | 50 (50, 51) | 14 (14, 15) | 50 (50, 51) |
| Average indexes per person | | | | | | |
|     Low severity | 0.75 (0.74, 0.75) | 0.75 (0.74, 0.76) | 1.15 (1.15, 1.16) | 0.61 (0.61, 0.62) | 0.96 (0.96, 0.97) | 0.61 (0.61, 0.62) |
|     High severity | 0.25 (0.25, 0.26) | 0.25 (0.24, 0.26) | 0.80 (0.79, 0.80) | 0.61 (0.61, 0.62) | 0.16 (0.16, 0.17) | 0.61 (0.61, 0.62) |
| *Scenario 2: severity does not impact visit cadence; effect modification by severity* | | | | | | |
| Number of people | 3124 (3090, 3152) | 1876 (1848, 1910) | 2526 (2506, 2550) | 2474 (2450, 2494) | 2526 (2506, 2550) | 2474 (2450, 2494) |
| Number of indexes | | | | | | |
|     Low severity | 2344 (2318, 2368) | 1407 (1380, 1433) | 2910 (2874, 2949) | 1520 (1490, 1540) | 2434 (2409, 2458) | 1520 (1490, 1540) |
|     High severity | 778 (764, 797) | 468 (456, 481) | 2016 (1987, 2036) | 1530 (1502, 1551) | 412 (400, 427) | 1530 (1502, 1551) |
| Percent high severity indexes | 25 (25, 25) | 25 (24, 26) | 41 (41, 41) | 50 (50, 51) | 14 (14, 15) | 50 (50, 51) |
| Average indexes per person | | | | | | |
|     Low severity | 0.75 (0.75, 0.75) | 0.75 (0.74, 0.76) | 1.15 (1.14, 1.16) | 0.61 (0.61, 0.62) | 0.96 (0.96, 0.97) | 0.61 (0.61, 0.62) |
|     High severity | 0.25 (0.25, 0.25) | 0.25 (0.24, 0.26) | 0.80 (0.79, 0.81) | 0.62 (0.61, 0.62) | 0.16 (0.16, 0.17) | 0.62 (0.61, 0.62) |
| *Scenario 3: severity impacts visit cadence; no effect modification* | | | | | | |
| Number of people | 3120 (3094, 3142) | 1880 (1858, 1906) | 1636 (1608, 1658) | 3364 (3342, 3392) | 1636 (1608, 1658) | 3364 (3342, 3392) |
| Number of indexes | | | | | | |
|     Low severity | 2342 (2317, 2371) | 1410 (1388, 1429) | 2151 (2120, 2187) | 2275 (2253, 2300) | 1614 (1587, 1635) | 2275 (2253, 2300) |
|     High severity | 775 (761, 794) | 470 (459, 485) | 1034 (1009, 1054) | 2500 (2479, 2524) | 582 (558, 596) | 2500 (2479, 2524) |
| Percent high severity indexes | 25 (24, 25) | 25 (24, 26) | 32 (32, 33) | 52 (52, 53) | 26 (26, 27) | 52 (52, 53) |
| Average indexes per person | | | | | | |
|     Low severity | 0.75 (0.75, 0.76) | 0.75 (0.74, 0.76) | 1.32 (1.31, 1.33) | 0.68 (0.67, 0.68) | 0.99 (0.98, 0.99) | 0.68 (0.67, 0.68) |
|     High severity | 0.25 (0.24, 0.25) | 0.25 (0.24, 0.26) | 0.63 (0.62, 0.64) | 0.74 (0.74, 0.75) | 0.35 (0.34, 0.36) | 0.74 (0.74, 0.75) |
| *Scenario 4: severity impacts visit cadence; effect modification by severity* | | | | | | |
| Number of people | 3127 (3108, 3150) | 1873 (1850, 1892) | 1641 (1617, 1659) | 3359 (3341, 3383) | 1641 (1617, 1659) | 3359 (3341, 3383) |
| Number of indexes | | | | | | |
|     Low severity | 2346 (2317, 2366) | 1402 (1381, 1426) | 2150 (2117, 2187) | 2270 (2250, 2294) | 1614 (1594, 1641) | 2270 (2250, 2294) |
|     High severity | 780 (767, 805) | 469 (457, 479) | 1044 (1021, 1060) | 2497 (2467, 2520) | 581 (569, 596) | 2497 (2467, 2520) |
| Percent high severity indexes | 25 (24, 26) | 25 (24, 26) | 33 (32, 33) | 52 (52, 53) | 27 (26, 27) | 52 (52, 53) |
| Average indexes per person | | | | | | |
|     Low severity | 0.75 (0.74, 0.76) | 0.75 (0.74, 0.76) | 1.31 (1.30, 1.32) | 0.68 (0.67, 0.68) | 0.99 (0.98, 0.99) | 0.68 (0.67, 0.68) |
|     High severity | 0.25 (0.24, 0.26) | 0.25 (0.24, 0.26) | 0.64 (0.62, 0.65) | 0.74 (0.73, 0.75) | 0.36 (0.35, 0.36) | 0.74 (0.73, 0.75) |

IQR = Interquartile range

**Table S4** Bias, root mean squared error (rMSE), and empirical standard error (ESE) estimates resulting from the crude log risk ratio, the log risk ratio standardized to the severity distribution in the entire cohort (ATE), and the log risk ratio standardized to the severity distribution in the treated subset of the cohort (ATT) across 5,000 simulated cohorts. Results are presented for the single point trial (SPT) and the sequential nested trial emulations with (eSNT-TD) and without (eSNT-CAL) a treatment decision requirement at the second visit.

| Population | | Crude[a] | | | | ATE | | | | ATT | | | |
|---|---|---|---|---|---|---|---|---|---|---|---|---|---|
| Analyzed | Target | RR | Bias[b] | rMSE | ESE | RR | Bias[b] | rMSE | ESE | RR | Bias | rMSE | ESE |
| *Scenario 1: severity does not impact treatment decision; no effect modification* | | | | | | | | | | | | | |
| SPT | SPT[c] | 0.70 | -0.00 | 0.07 | 0.53 | 0.70 | -0.00 | 0.07 | 0.52 | 0.70 | -0.00 | 0.07 | 0.52 |
| eSNT-CAL | eSNT-CAL[c] | 0.78 | 0.11 | 0.12 | 0.43 | 0.70 | -0.00 | 0.06 | 0.45 | 0.70 | -0.00 | 0.06 | 0.43 |
|  | SPT |  |  |  |  | 0.70 | -0.00 | 0.07 | 0.52 | 0.70 | -0.00 | 0.07 | 0.52 |
| eSNT-TD | eSNT-TD[c] | 0.89 | 0.24 | 0.25 | 0.47 | 0.70 | -0.00 | 0.08 | 0.54 | 0.70 | -0.00 | 0.08 | 0.57 |
|  | SPT |  |  |  |  | 0.70 | -0.00 | 0.08 | 0.56 | 0.70 | -0.00 | 0.08 | 0.56 |
| *Scenario 2: severity does not treatment decision; effect modification* | | | | | | | | | | | | | |
| SPT | SPT[c] | 0.70 | -0.00 | 0.07 | 0.51 | 0.70 | -0.00 | 0.07 | 0.51 | 0.70 | -0.00 | 0.07 | 0.51 |
| eSNT-CAL | eSNT-CAL[c] | 0.91 | 0.25 | 0.26 | 0.41 | 0.77 | 0.09 | 0.10 | 0.41 | 0.81 | 0.14 | 0.15 | 0.40 |
|  | SPT |  |  |  |  | 0.71 | -0.00 | 0.07 | 0.49 | 0.71 | -0.00 | 0.07 | 0.49 |
| eSNT-TD | eSNT-TD[c] | 1.04 | 0.39 | 0.39 | 0.44 | 0.73 | 0.04 | 0.08 | 0.50 | 0.81 | 0.14 | 0.16 | 0.54 |
|  | SPT |  |  |  |  | 0.71 | -0.00 | 0.07 | 0.52 | 0.71 | -0.00 | 0.07 | 0.52 |
| *Scenario 3: severity impacts treatment decision; no effect modification* | | | | | | | | | | | | | |
| SPT | SPT[c] | 0.70 | -0.00 | 0.07 | 0.52 | 0.70 | -0.00 | 0.07 | 0.52 | 0.70 | -0.00 | 0.07 | 0.52 |
| eSNT-CAL | eSNT-CAL[c] | 0.88 | 0.23 | 0.23 | 0.45 | 0.70 | 0.00 | 0.07 | 0.48 | 0.70 | 0.00 | 0.07 | 0.47 |
|  | SPT |  |  |  |  | 0.70 | 0.00 | 0.08 | 0.56 | 0.70 | 0.00 | 0.08 | 0.56 |
| eSNT-TD | eSNT-TD[c] | 0.91 | 0.26 | 0.27 | 0.48 | 0.70 | 0.00 | 0.07 | 0.52 | 0.70 | 0.00 | 0.08 | 0.54 |
|  | SPT |  |  |  |  | 0.70 | 0.00 | 0.08 | 0.58 | 0.70 | 0.00 | 0.08 | 0.58 |
| *Scenario 4: severity impacts treatment decision; effect modification* | | | | | | | | | | | | | |
| SPT | SPT[c] | 0.70 | -0.00 | 0.07 | 0.51 | 0.70 | -0.00 | 0.07 | 0.51 | 0.70 | -0.00 | 0.07 | 0.51 |
| eSNT-CAL | eSNT-CAL[c] | 1.05 | 0.40 | 0.40 | 0.44 | 0.77 | 0.09 | 0.11 | 0.46 | 0.84 | 0.17 | 0.18 | 0.46 |
|  | SPT |  |  |  |  | 0.71 | -0.00 | 0.08 | 0.54 | 0.71 | -0.00 | 0.08 | 0.54 |
| eSNT-TD | eSNT-TD[c] | 1.08 | 0.43 | 0.43 | 0.47 | 0.77 | 0.09 | 0.11 | 0.51 | 0.84 | 0.17 | 0.19 | 0.54 |
|  | SPT |  |  |  |  | 0.71 | -0.00 | 0.08 | 0.57 | 0.71 | -0.00 | 0.08 | 0.57 |

rMSE = root mean-squared error; ESE = empirical standard error
[a] The "crude" estimates were adjusted for informative right-censoring but not for confounding due to severity.
[b] Monte-Carlo standard error for the bias ranged from 0.005 to 0.008.
[c] Indicates the target population is the same as the analyzed population